\documentclass[aps,prb,reprint,showkeys]{revtex4-1}

\usepackage{subcaption}
\usepackage{bm,graphicx,tabularx,array,booktabs,dcolumn,xcolor,microtype,multirow,amscd,amsmath,amssymb,amsfonts,physics,siunitx,enumitem,upgreek}
\usepackage[version=4]{mhchem}
\usepackage[normalem]{ulem}
\usepackage{mleftright}
\usepackage{wrapfig}

\usepackage[utf8]{inputenc}
\usepackage[T1]{fontenc}

\AtBeginDocument{\nocite{achemso-control}}

\usepackage{tikz}
\usetikzlibrary{arrows.meta,positioning,shapes.misc,arrows}

\newcommand{\cA}{\mathcal{A}}
\newcommand{\cC}{\mathcal{C}}
\newcommand{\cP}{\mathcal{P}}
\newcommand{\cQ}{\mathcal{Q}}
\newcommand{\cR}{\mathcal{R}}

\newcommand{\cV}{\mathcal{V}}
\newcommand{\cW}{\mathcal{W}}

\newcommand{\MO}{\psi}

\newcommand{\Nb}{n}
\newcommand{\Ns}{N_\text{s}}
\newcommand{\No}{N_\text{o}}

\definecolor{hughgreen}{RGB}{100, 0, 140}

\newcommand{\suppI}{Supplementary Material}

\allowdisplaybreaks

\usepackage[
	colorlinks=true,
    citecolor=blue,
    linkcolor=blue,
    filecolor=blue,      
    urlcolor=blue,	
    breaklinks=true
	]{hyperref}
\newcommand{\UCL}{Department of Chemistry, University College London, London, WC1H 0AJ, U.K.}

\begin{document}	

\title{Geometric direct minimization for low-spin restricted open-shell Hartree--Fock}

\date{\today}
\author{Hugh G.~A.~\surname{Burton}}
\email{h.burton@ucl.ac.uk}
\affiliation{\UCL}

\begin{abstract}
\begin{center}\textbf{Abstract}\end{center}\vspace{-1em}
It has recently been shown that configuration state functions (CSF) with local orbitals can provide a 
compact reference state for low-spin open-shell electronic structures, such as antiferromagnetic states.
However, optimizing a low-spin configuration using self-consistent field (SCF) theory has been 
a long-standing challenge, since each orbital must be an eigenfunction of a different Fock operator.
Here, I introduce a low-spin restricted open-shell Hartree--Fock (ROHF) algorithm to optimize any 
CSF at mean-field cost.
This algorithm employs quasi-Newton Riemannian optimization on the orbital constraint manifold 
to provide robust convergence, 
extending the Geometric Direct Minimization approach to open-shell electronic structures with
arbitrary genealogical spin coupling. 
Numerical calculations on transition metal aquo complexes show improved convergence over 
existing methodology, while the possibility of local CSF energy minima is demonstrated for iron-sulfur complexes.
Finally, open-shell CSFs with different spin coupling patterns are used to qualitatively
study the singlet ground state in polyacenes, revealing the onset of polyradical character for
increasing chain length.
\end{abstract}

\maketitle
\raggedbottom

\section{Introduction}
\label{sec:intro}

Open-shell electronic configurations underpin quantum phenomena such as the magnetic 
properties of transition metal complexes,\cite{Malrieu2013}
the spin-state energetics of radicals,\cite{Krylov2017}
 and molecular excited states.\cite{Gonzalez2012}
However, theoretically characterising open-shell states is complicated 
due to the near degeneracy of configurations with complex spin alignment, such as 
ferromagnetic and antiferromagnetic states.
Developing  efficient methods that accurately predict low-lying energy states of open-shell 
systems, while retaining conceptual understanding, remains a major challenge.


Wavefunction predictions of open-shell states require multiconfigurational 
approximations that account for the ``static correlation'' associated with nearly 
degenerate configurations.
The complete active space self-consistent field (CASSCF) approach is the most common, whereby 
a full configuration interaction (FCI) is constructed within a set of active orbitals that 
are optimized simultaneously.\cite{Roos1980a,Roos1980b}
However, CASSCF calculations are notoriously challenging because:
the computational cost scales exponentially with the size of the active space;
results are sensitive to the choice of active orbitals;\cite{Cabrero2002}
and the numerical optimization can be poorly conditioned,\cite{Yeager1979,Yeager1980a,Jorgensen1981,Yeager1982,Olsen1983,Werner1985} 
with many possible stationary points.\cite{Marie2023,Saade2024}
Furthermore, large active spaces are required to accurately compute low-lying states in polynuclear  
transition metal complexes
or extended conjugated molecules.\cite{Sharma2014,Hachmann2007}
These calculations quickly become intractable for exact diagonalization and 
rely on approximate solvers\cite{Eriksen2020} including FCI quantum Monte Carlo (FCIQMC),\cite{Booth2009}
density matrix renormalization group (DMRG),\cite{White1992,Chan2002,Hachmann2007} 
or selected CI.\cite{Huron1973,Holmes2016,Giner2013,Evangelista2014,Tubman2016}
Even then, computing ``dynamic correlation'' on top of a CASSCF wavefunction remains a 
formidable challenge.

The complexity of CASSCF theory raises the question of whether alternative 
single-reference methods can be designed to encode the dominant static correlation and 
spin coupling without the need for an active space.
It is well known that a single Slater determinant can provide a good approximation to 
high-spin systems (with unpaired electrons all spin aligned)
but is inadequate for low-spin cases.\cite{Krylov2017}
However, it has only recently been discovered that many antiferromagnetic low-spin states
can be accurately represented by a small number of configuration state functions\cite{HelgakerBook} (CSF), 
sometimes only one.\cite{Manni2020,Manni2021a,Manni2021b,Dobrautz2022}
This approach relies on localized molecular orbitals (MOs)
and an appropriate orbital ordering that combines local ferromagnetic coupling with long-range
antiferromagnetic coupling.\cite{Manni2020,Dobrautz2022,Izsak2023,MartiDafcik2024}
Using an optimal CSF basis can significantly reduce the multireference character of the
wavefunction and provide a sparser representation of the Hilbert space compared to an 
RHF-based determinant basis.\cite{Manni2020,Dobrautz2022,Iszak2023}
These properties have been exploited to accelerate the convergence 
of spin-adapted FCI solvers, such as GUGA-FCIQMC\cite{Dobrautz2019,Dobrautz2021}
or selected CI,\cite{Chilkuri2021a,Chilkuri2021b}
and to define accurate initial states for future quantum computing algorithms.\cite{MartiDafcik2024,MartiDafcik2025}


In practice, a major challenge to using single CSF reference states for open-shell correlation theory is 
finding the optimal MOs through an initial Hartree--Fock (HF) calculation.
While the HF equations are straightforward to solve for a closed-shell determinant,\cite{Roothaan1951,Hall1951}  
this becomes much harder for a low-spin open-shell configuration since each optimal spatial orbital $\MO_i$
is an eigenfunction of a different Fock operator $\hat{f}_i $, 
satisfying  $\hat{f}_i \ket{\MO_i} = \epsilon_i \ket{\MO_i}$.\cite{Roothaan1960}
(Note that orbitals experiencing the same Fock operator are said to occupy the same ``shell''.)
While several algorithms to solve the Roothaan--Hall equations for restricted open-shell HF (ROHF) 
have been developed,\cite{Roothaan1960,Segal1970,Peters1972,Binkley1974,Guest1974,Edwards1987,Tsuchimochi2010} 
their generalization to low-spin 
CSFs with arbitrary spin coupling was only recently achieved by Neese and co-workers.\cite{Gouveia2024}
However, SCF algorithms based on Fock diagonalization can be difficult to converge in the case of near degeneracies and are not guaranteed to converge to an energy minimum.

The aim of this work is to develop a quasi-Newton direct minimization algorithm for arbitrary low-spin CSF states that 
provides robust convergence to an energy minimum.
Quasi-Newton optimization techniques can significantly improve SCF convergence in challenging cases,
and have been widely adopted for HF and multiconfigurational SCF calculations.\cite{Backsay1981,Backsay1982,Douady1980,Fischer1992,Chaban1997,Olsen1983,Yeager1979,Yeager1980a,Werner1985,Jorgensen1981,Yeager1982,Neese2000,Voorhis2002,Dunietz2002,Kreplin2019,Kreplin2020,Slattery2024}
A particularly successful approach is Geometric Direct Minimization (GDM),\cite{Voorhis2002} which takes 
into account the Riemannian geometry of the orthonormal MO coefficients for a single Slater determinant.
While GDM has previously been extended to high-spin ROHF calculations,\cite{Dunietz2002,Vidal2024} 
here I introduce a general formulation for a single CSF with arbitrary spin coupling.
The resulting ``CSF-GDM'' approach provides robust energy minimization for arbitrary low-spin CSFs, 
avoiding the need to handle a different Fock operator for each shell.

Developing the CSF-GDM approach provides two opportunities to further study the utility 
of CSF-based ROHF theory that are outlined below.

Firstly, we currently have limited knowledge about the properties of optimal CSF solutions. 
For example, does the physically intuitive orbital localization and ordering actually exist as a  minimum of the CSF energy?
Furthermore, it is known that unrestricted HF can yield many local minima
for open-shell systems, associated with localising the unpaired 
electrons and breaking spin symmetry,\cite{Slater1951,Fukutome1971,Davidson1983,Li2009a,Kowalski1998,Lee2019,Shee2021,Toth2016,Thom2008,Thompson2018,Burton2021b,Burton2021a,Saade2024}
and it is important to assess whether CSF-based ROHF is also susceptible to multiple minima. 
The CSF-GDM algorithm allows the electronic energy landscape\cite{Burton2021a,Burton2022a} of CSF-based ROHF theory
to be systematically investigated by ensuring 
that calculations initialized with randomly perturbed orbital coefficients converge to a local minimum.
Here, I test this approach for different spin states in model iron-sulfur clusters,  
revealing that many local minima can exist and that solutions with unpaired electrons localized in \ce{Fe} 3d orbitals 
(which might be predicted from chemical intuition) are not necessarily local minima for all CSF spin states.

Secondly,  broken-spin Kohn--Sham Density Functional Theory (KS-DFT) 
is currently the method of choice to qualitatively understand the electron localization in open-shell states.
However, the presence of broken spin symmetry is not a perfect indicator of open-shell character since ``artificial'' 
symmetry breaking is well-documented in molecules that would normally be considered to have a closed-shell ground
state.\cite{Davidson1983,Cohen2001,Lee2019,Pillai2025}
I show that CSF-based ROHF theory can provide qualitative insights into electron localization by comparing 
the energies of open-shell CSFs with different numbers of unpaired electrons, retaining both a mean-field 
computational cost and conserving spin symmetry.
Applying this approach to the singlet ground state in polyacenes 
reveals the onset of polyradical character as the number of rings increases, 
confirming previous predictions using broken-spin KS-DFT.\cite{Trinquier2018}


The remainder of this work is structured as follows.
Section~\ref{sec:rohf} describes the differential geometry of the ROHF wave function and energy.
Although many of these expressions have been derived elsewhere, a comprehensive description is provided for reference and completeness.
The CSF-GDM algorithm is then derived for arbitrary genealogical spin coupling in Section~\ref{sec:GDM}, with 
computational details in Section~\ref{sec:comp}.
Numerical results detailing the convergence performance of CSF-GDM, the multiple solutions for 
the iron-sulfur complexes   \ce{[Fe(SCH3)_4]^-} and  \ce{[Fe_2S_2(SCH3)_4]^2-}, 
and the open-shell character of polyacenes are described in Section~\ref{sec:results}.
The primary conclusions and outlook for future work are summarized in Section~\ref{sec:conclusions}.

\section{Differential geometry of the ROHF energy}
\label{sec:rohf}

\subsection{Definition of a configuration state function}

A CSF corresponds to a spin-adapted  linear combination of 
Slater determinants that is an eigenstate of both the spatial orbital number operator $\hat{n}_p$ and $\hat{S}^2$.\cite{HelgakerBook}
Here, doubly-occupied closed-shell orbitals are indexed $i,j,k$, singly-occupied open-shell orbitals are indexed $v,w,x$, and unoccupied orbitals are index $a,b,c$. 
Arbitrary orbitals are indexed $p,q,r$.
The genealogical coupling scheme is the most common method to build a CSF, 
whereby open-shell electrons are sequentially
coupled while maintaining an eigenstate of $\hat{S}^2$.
A particular CSF spin coupling is specified by a vector $\bm{t}$, where $t_i$ denotes the change in the total spin $S$
by coupling the $i$-th open-shell electron to the $(i-1)$ previous open-shell electrons.
For example, $\bm{t} = (+\frac{1}{2},-\frac{1}{2})$, denoted [+\textminus] for convenience, 
corresponds to the open-shell singlet ($S=0$)
\begin{equation}
\ket{\Psi} = \frac{1}{\sqrt{2}} \ket{\MO_1 \MO_2} \qty(\ket{\alpha \beta} - \ket{\beta \alpha}),
\end{equation}
while [++] denotes the high-spin triplet state ($S=1$)
\begin{equation}
\ket{\Psi} = \ket{\MO_1 \MO_2} \ket{\alpha \alpha}.
\end{equation}
The length of $\bm{t}$ defines the number of open-shell electrons $\No$, with the remaining core orbitals doubly occupied,
and the total spin $S$ is the sum $S=\sum_{i=1}^{\No} t_i$.

\begin{figure}[htb]
\includegraphics[width=0.9\linewidth]{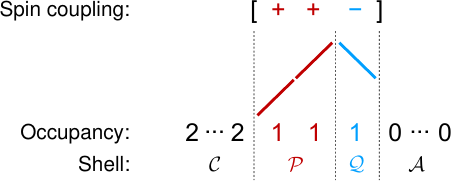}
\caption{The orbital ``shells'' for the [++\textminus] CSF.
Each partially-occupied shell is a straight-line segment in the genealogical branching diagram.\cite{HelgakerBook}
The doubly occupied and unoccupied orbitals  form shells $\cC$ and $\cA$, respectively.}
\label{fig:shells}
\end{figure}

The unpaired electrons can be grouped into sets of sequential $+$'s or $-$'s in the spin coupling vector. 
These groups are historically known as ``shells'' in ROHF theory because the electrons within each group 
experience the same effective 1-body Hamiltonian.\cite{Roothaan1960}
For example, the singlet [+\textminus] contains two open shells, [+] and [\textminus], while the triplet 
[++] has only one open shell.
Transformations that mix orbitals in the same shell leave the wavefunction unchanged, except for a global phase,
and the doubly-occupied orbitals and unoccupied virtual orbitals are considered as additional shells. 
Here, different  shells (including the doubly occupied and unoccupied shells) 
are labelled with the calligraphic indices $\cP, \cQ$ etc and 
the cardinality $D_{\cP} \equiv\abs{\cP}$ is the number of spatial orbitals in shell $\cP$.
The shell structure of the doublet configuration [++\textminus] with the doubly occupied shell $\cC$, two 
singlet occupied shells $\cP \equiv$ [++] and $\cQ \equiv [-]$ and the unoccupied shell $\cA$ is illustrated 
schematically in Fig.~\ref{fig:shells}.
Crucially, a CSF is not invariant to mixing orbitals in different shells, so the wavefunction depends on 
the assignment of spatial orbitals to each shell, often referred to as the orbital ordering.\cite{Manni2020,Dobrautz2022}

In general, there are several spin coupling vectors for each $S$, 
which form an orthonormal CSF basis for a fixed spatial orbital ordering.
Allowed spin coupling vectors must have
non-negative partial spins ($\sum_{j=1}^i t_j \geq 0$ for all $i$) with $t_1 = +\frac{1}{2}$.
The Clebsch--Gordon coupling coefficients allow any CSF to be expanded as a sum of 
Slater determinants with the same spatial orbital occupation and spin projection $m_s = S$.
In practice, defining a CSF through genealogical coupling allows matrix elements to
be efficiently evaluated using the unitary group approach\cite{Paldus1974,Paldus1976a,Paldus1976b} (UGA) 
and its graphical extension\cite{Shavitt1977} (GUGA).

\subsection{Energy of a configuration state function}
The energy of a wave function $\ket{\Psi}$ depends on the (partially) occupied 
MOs $\ket*{\MO_p}$ as 
\begin{equation}
E = \sum_{pq} h_{pq} \gamma_{pq} + \frac{1}{2}\sum_{pqrs} \Gamma_{pqrs} \expval{pq|rs},
\label{eq:energy}
\end{equation}
where $h_{pq}$ denote the one-electron integrals, $\expval{pq|rs}$ are the two-electron repulsion integrals,
and $\gamma_{pq}$ and $\Gamma_{pqrs}$ are the one- and two-electron reduced density matrices (RDMs) 
in the spatial MO basis.
By convention, the 1- and 2-RDMs are defined in second quantization as
\begin{subequations}
\begin{align}
\gamma_{pq} &= \mel{\Psi}{\hat{E}_{pq}}{\Psi} 
\\
\Gamma_{pqrs} &= \mel{\Psi}{\hat{e}_{pr,qs}}{\Psi}.
\end{align}
\end{subequations}
where the singlet excitation operator is defined as
$\hat{E}_{pq} = \hat{a}^{\dagger}_{p \uparrow} \hat{a}^{\vphantom{\dagger}}_{q \uparrow} + \hat{a}^{\dagger}_{p \downarrow} \hat{a}^{\vphantom{\dagger}}_{q \downarrow}$,
and $\hat{e}_{pr,qs} = \hat{E}_{pr}\hat{E}_{qs}-\delta_{rq}\hat{E}_{ps}$.\cite{Paldus1974,HelgakerBook}

The spatial orbital occupancy for a single CSF must be conserved for a density matrix element to be non-zero. 
Therefore, the only non-zero terms are
\begin{subequations}
\begin{align}
\gamma_{pq} &= \delta_{pq} n_p,
\\
\Gamma_{pqpq} &= n_p n_q - \delta_{pq} n_p 
\\
\Gamma_{pqqp} &= \mel{\Psi}{\hat{E}_{pq}\hat{E}_{qp}}{\Psi} - n_p,
\end{align}
\label{eq:rdm}
\end{subequations}
where $n_p$ is the occupation number of spatial orbital $\MO_p$.
By defining the constants $a_{pq}$ and $b_{pq}$ as
\begin{align}
a_{pq} \equiv \Gamma_{pqpq} \quad\text{and}\quad
b_{pq} \equiv \Gamma_{pqqp} - \delta_{pq} \Gamma_{pqpq},
\label{eq:vectorconstants}
\end{align}
the CSF energy is then given in a similar form to Roothaan's ROHF expression as\cite{Roothaan1960}
\begin{equation}
E =  \sum_{p} h_{pp} n_p + \frac{1}{2}\sum_{pq} \qty(a_{pq} \expval{pq|pq} + b_{pq} \expval{pq|qp}),
\label{eq:Eab}
\end{equation}
where the term $- \delta_{pq} \Gamma_{pqpq}$ in the definition of $b_{pq}$
avoids double counting of the two-electron interactions for $p=q$.
The energy difference between CSFs with the same orbital occupation comes from only
the exchange contribution $\mel{\Psi}{\hat{E}_{pq}\hat{E}_{qp}}{\Psi}$.
Since $\expval{pq|pq} = \expval{pq|qp}$ when $p=q$, Eq.~\eqref{eq:Eab} is invariant to an equal and opposite shift 
in the diagonal constants $a_{pp}$ and $b_{pp}$ as long as $a_{pp} + b_{pp}$ remains unchanged.
Using $\mel{\Psi}{\hat{E}_{pp}\hat{E}_{pp}}{\Psi} = n_p n_p$ then yields
sufficient expressions for $a_{pq}$ and $b_{pq}$ as
\begin{subequations}
\begin{align}
a_{pq} &\equiv n_p n_q,
\\
b_{pq} &\equiv (1-\delta_{pq})\mel{\Psi}{\hat{E}_{pq}\hat{E}_{qp}}{\Psi} - n_p.
\label{eq:b1}
\end{align}
\label{eq:vcc_csf}
\end{subequations}
Explicit constants for the closed-shell orbitals $i,j$, and open-shell orbitals $v,w$ are found to be
\begin{subequations}
\begin{alignat}{2}
a_{ij} &= 4, \quad &&b_{ij} = - 2 \\
a_{iv} &= 2, \quad &&b_{iv} = - 1 \\
a_{vw} &= 1, \quad &&b_{vw} = \mel{\Psi}{\hat{E}_{vw}\hat{E}_{wv}}{\Psi} - 1 - \delta_{vw},
\label{eq:b2}
\end{alignat}
\end{subequations}
where $a_{vi} = a_{iv}$ and $b_{vi} = b_{iv}$.
Here, $\mel{\Psi}{\hat{E}_{ww}\hat{E}_{ww}}{\Psi} = 1$ is used to simplify Eq.~\eqref{eq:b1} into Eq.~\eqref{eq:b2}.
These expressions are equivalent to Eqs.~(7--9) in Ref.~\citenum{Gouveia2024}, although the definition for 
the open-shell coupling constants differs by a factor of 2.
Since all open-shell orbitals $v$ in shell $\cV$ and $w$ in shell $\cW$ share the same 
matrix element $\mel{\Psi}{\hat{E}_{vw}\hat{E}_{wv}}{\Psi} = \mel{\Psi}{\hat{e}_{vw,wv}}{\Psi} + \delta_{vw}$, the constants $b_{\cV \cW} \equiv b_{vw}$ can be 
evaluated once for each pair of shells. \cite{Edwards1987,Gouveia2024}
Following Ref.~\citenum{Gouveia2024}, the $b_{vw}$ constants 
can be found by using the GUGA method\cite{Shavitt1977} 
to derive the exchange terms $\mel{\Psi}{\hat{e}_{vw,wv}}{\Psi}$ for open-shell indices $v$ and $w$. 
An excellent derivation of these terms for a single CSF can be found in 
Appendix~B of Ref.~\citenum{Dobrautz2019}.

\subsection{Analytic gradients and second derivatives}
The spatial MOs are parametrized in terms of $\Nb$ linearly independent atomic orbital (AO) 
basis functions $\ket*{\chi_\mu}$ as 
\begin{equation}
\ket{\MO_p} = \sum_{\mu=1}^{\Nb} \ket{\chi_\mu} C^{\mu \cdot}_{\cdot p},
\end{equation}
where nonorthogonal tensor notation is used.\cite{HeadGordon1994}
The energy [Eq.~\eqref{eq:Eab}] can then be minimized with respect to the orbital 
coefficients $C^{\mu \cdot}_{\cdot p}$ under the orthonormality constraint
\begin{equation}
\sum_{\mu \nu=1}^{\Nb} C^{\cdot \mu}_{p \cdot} \braket*{\chi_\mu}{\chi_\nu} C^{\nu \cdot}_{\cdot q} = \delta_{pq},
\end{equation}
where the orbital coefficients are chosen to be real, $C^{\nu \cdot}_{\cdot q} \in \mathbb{R}$.
To satisfy orthonormality, the matrix of orbital coefficients $\bm{C}$ is constrained to a Riemannian manifold corresponding to
the orthogonal group $\mathrm{O}(\Nb)$.\cite{Edelman1998,Voorhis2002}
However, since transformations between orbitals in the same shell leave the energy unchanged, 
only transformations between different shells should be considered as optimization parameters.
The CSF optimization manifold then corresponds to the quotient space\cite{Edelman1998}
\begin{equation}
\frac{\mathrm{O}(\Nb)}{\mathrm{O}(D_{\cP}) \times \cdots \times \mathrm{O}(D_{\cQ})}
\label{eq:flag}
\end{equation}
where each subgroup $\mathrm{O}(D_{\cP})$ represents the invariance to orthogonal transformations among the $D_{\cP}$ orbitals in shell $\cP$.
This structure corresponds to a flag manifold, as described for high-spin ROHF in Ref.~\citenum{Vidal2024},
although Eq.~\eqref{eq:flag} provides the generalization to arbitrary low-spin open-shell configurations 
with more than three invariant subspaces.

The non-redundant variations of an initial CSF $\ket{\Psi_0}$ with orbital coefficients $\bm{C}_0$ 
can be parametrized using an exponential transformation as\cite{Douady1980}
\begin{equation}
\ket{\Psi (\bm{\kappa})} = \exp(\hat{\kappa}) \ket{\Psi_0},
\label{eq:psi_kappa}
\end{equation}
where $\hat{\kappa}$ is an anti-Hermitian one-body operator that only couples electrons in different shells as
\begin{equation}
\hat{\kappa} = \sum_{\cP\neq \cQ} \sum_{p \in \cP} \sum_{q \in \cQ} \kappa_{pq} \qty(\hat{E}_{pq} - \hat{E}_{qp}).
\end{equation}
The orbital rotation step $\bm{\kappa}$ is an anti-Hermitian $\Nb \times \Nb$ matrix with an 
off-diagonal block structure, e.g., for  four shells
\begin{equation}
\bm{\kappa} =
\begin{pmatrix}
\bm{0}           & -\bm{\kappa}_{21}^\dagger & -\bm{\kappa}_{31}^\dagger & -\bm{\kappa}_{41}^\dagger \\
\bm{\kappa}_{21} & \bm{0}                    & -\bm{\kappa}_{32}^\dagger & -\bm{\kappa}_{42}^\dagger \\
\bm{\kappa}_{31} & \bm{\kappa}_{32}          & \bm{0}                    & -\bm{\kappa}_{43}^\dagger \\
\bm{\kappa}_{41} & \bm{\kappa}_{42}          & \bm{\kappa}_{43}          & \bm{0}
\end{pmatrix}.
\label{eq:tangent_form}
\end{equation}
Since a one-body transformation corresponds to an orbital rotation,
Eq.~\eqref{eq:psi_kappa} is equivalent to the transformation 
\begin{equation}
\bm{C}(\bm{\kappa}) = \bm{C}_0 \exp(\bm{\kappa}).
\label{eq:MOupdate}
\end{equation} 
In practice, the orbital coefficients are updated on each iteration $k$ as
\begin{equation}
\bm{C}_{k+1} = \bm{C}_k \exp(\bm{\kappa}),
\label{eq:MOupdate_step}
\end{equation}
such that the step $\bm{\kappa}$ is always expressed in the local MO basis.\cite{Voorhis2002}
The advantage of parametrising a CSF on a continuous manifold is that the assignment of orbitals 
to each shell is controlled 
by the ordering of columns in $\bm{C}$, providing more robust convergence compared to to Fock 
diagonalization schemes where orbitals must be allocated to shells on every iteration.\cite{Edwards1987,Peters1972,Roothaan1960}

Analytic gradients and second-derivatives
can now be derived as a special case of multi-configurational SCF,\cite{Chaban1997,Dalgaard1978,Dalgaard1979,Douady1980} 
as reviewed extensively in Ref.~\citenum{HelgakerBook}.
The gradient components are 
\begin{equation}
g_{pq} \equiv \left.\frac{\partial E}{\partial \kappa_{pq}}\right|_{\bm{\kappa}=\bm{0}} = 2\,\qty(F_{pq} - F_{qp}),
\end{equation}
where $F_{pq}$ are elements of the (non-symmetric) generalized Fock matrix, defined in the MO basis as\cite{HelgakerBook} 
\begin{equation}
F_{pq} = \sum_{r=1}^\Nb \gamma_{pr} h_{rq} + \sum_{rst=1}^\Nb \Gamma_{prst} \expval{st|qr}.
\label{eq:genfock}
\end{equation}
Note that $F_{aq}=0$ if the first index corresponds to an orbital in the unoccupied shell.
The generalized Fock matrix element for cases where the first index $p$ corresponds to an MO in shell $\cP$ is given by
\begin{equation}
F_{pq}^\cP = n_\cP \qty(h_{pq} + J_{pq}) + K^\cP_{pq},
\label{eq:genfock_JK}
\end{equation}
where the superscript $\cP$ indicates that the matrix element is computed using the exchange operator 
$K_{pq}^\cP$ experienced by shell $\cP$.
The total Coulomb operator $J_{pq}$ and the shell exchange operator $K_{pq}^\cP$ are defined as
\begin{subequations}
\begin{align}
J_{pq} &= \sum_{r} n_r \expval{pr|qr}, \\
K_{pq}^\cP &= \sum_{\cR}  b_{\cP \cR} \sum_{r\in\cR} \expval{pr|rq}.
\end{align}
\label{eq:JKmo}
\end{subequations}
Crucially, these Coulomb and exchange matrices can be evaluated using standard JK-builds in the AO basis, 
allowing the gradient to be obtained with $\mathcal{O}(\Ns\, \Nb^4)$ scaling, 
where $\Ns$ is the number of shells.
The gradient is then given explicitly as
\begin{equation}
g_{pq} = 2\,\qty(F_{pq}^\cP - F_{qp}^\cQ).
\end{equation}

The Hessian matrix of second derivatives,\cite{HelgakerBook} defined as
\begin{equation}
Q_{pq,rs} \equiv \left.\frac{\partial^2 E}{\partial \kappa_{pq} \partial \kappa_{rs}}\right|_{\bm{\kappa}=\bm{0}}
\end{equation}
can be obtained in analytic form for a single CSF as
\begin{equation}
\begin{split}
Q_{pq,rs} &=P_{pq}P_{rs} \Big[2\delta_{pr}F^\cP_{qs} - \delta_{qs}(F^\cP_{pr} + F^\cR_{rp})
\\
&+2\big( 2a_{pr}\expval{qr|ps} + b_{pr} \qty(\expval{qp|rs} + \expval{qr|sp}) \big) \Big].
\end{split}
\label{eq:Qpqrs}
\end{equation}
where the operator $P_{pq} = 1 - (pq)$ introduces an antisymmetric permutation of the indices $p$ and $q$.
A full derivation of Eq.~\eqref{eq:Qpqrs} is provided in Appendix~\ref{sec:hessian}.
Crucially, this expression can be used to precondition the optimization and accelerate convergence
(Section~\ref{sec:preconditioner}).

\section{Geometric Direct Minimization for an arbitrary open-shell CSF}
\label{sec:GDM}
Section~\ref{sec:rohf} introduced the necessary pre-requisites to develop a Riemannian optimization algorithm for a CSF with arbitrary geneaological spin coupling.
In contrast to optimization in flat Euclidean spaces, Riemannian optimization on a smooth manifold
takes into account the curvature of the manifold and changes in the tangent space at different points.\cite{Edelman1998,AbsilBook} 
The manifold curvature means that tangent vectors must be carefully translated between points 
to ensure that they remain in tangent space, using a process known as parallel transport (Fig.~\ref{fig:geom}).
Riemannian optimization accounts for this parallel transport to provide robust convergence in a curved space.
Riemannian optimization based on the L-BFGS algorithm\cite{Broyden1970,Fletcher1970,Goldfarb1970,Shanno1970}
 has previously been applied to single determinant SCF theory in the GDM algorithm,\cite{Voorhis2002,Dunietz2002}
and has recently been extended to high-spin ROHF and CASSCF theory.\cite{Vidal2024}
Here, I extend GDM to the case of a low-spin CSF, providing robust 
optimization for any geneaological spin coupling and an arbitrary number of open shells.

\subsection{Quasi-Newton L-BFGS optimization}
\label{sec:BFGS}
Quasi-Newton methods use the local gradient $\bm{g}$ and steps $\bm{s}$ to estimate
the inverse Hessian matrix and provide approximate second-order optimization of a multivariate function $f(\bm{x})$.\cite{NocedalBook}
On each iteration $k$, an approximation to the inverse Hessian matrix $\bm{B}_{k+1}$ is constructed and a downhill 
step $\bm{s}_{k+1}$ is identified using the Newton--Raphson formula
\begin{equation}
\bm{s}_{k+1} = - \bm{B}_{k+1} \, \bm{g}_{k+1}.
\label{eq:NRstep}
\end{equation}
In the BFGS algorithm,  $\bm{B}_k$ is defined recursively as\cite{Broyden1970,Fletcher1970,Goldfarb1970,Shanno1970}
\begin{equation}
\bm{B}_{k+1} = 
(\bm{I} - \rho_k \bm{s}_k^{\vphantom{\dagger}} \bm{y}_k^\dagger) 
\bm{B}_k 
(\bm{I} - \rho_k \bm{s}_k^{\vphantom{\dagger}} \bm{y}_k^\dagger) + \rho_k \bm{s}^{\vphantom{\dagger}}_k \bm{s}_k^\dagger,
\label{eq:BFGSupdate}
\end{equation}
where the standard definitions are
\begin{align}
\bm{y}_k = \bm{g}_{k+1} - \bm{g}_k, \quad
\bm{s}_k = \bm{x}_{k+1} - \bm{x}_k, \label{eq:BFGSsk} \quad 
\rho_k = \frac{1}{\bm{y}_k^\dagger\bm{s}_k^{\vphantom{\dagger}}}.
\end{align}
The definition of Eq.~\eqref{eq:BFGSupdate} maintains a positive definitive approximate inverse Hessian, 
which ensures downhill optimization steps.
In the limited-memory  L-BFGS variant,\cite{Nocedal1980} $\bm{B}_k$ is constructed using gradients and steps from only the 
$m$ most recent iterations, where typically $m \sim 10$--20.\cite{NocedalBook}
Suitable choices of the initial inverse Hessian approximation $\bm{B}^0_{k+1}$ can significantly accelerate convergence by 
preconditioning the optimization (see Section~\ref{sec:preconditioner}).

For  CSF  optimization, the exponential MO parametrization means 
that analytic gradients can only be computed at $\bm{\kappa} = \bm{0}$.\cite{Douady1980}
Therefore, instead of using global coordinates, the origin is updated on each iteration 
using Eq.~\eqref{eq:MOupdate_step} with the orbital rotation step $\bm{\kappa} = \bm{s}_{k+1}$,
before the new gradient is computed in local coordinates corresponding to local MO--MO transformations.
The local orbital rotation steps $\bm{\kappa}$ are then directly used to define $\bm{s}_k$ in the L-BFGS update.
An approximate line search using the Wolfe conditions\cite{Wolfe1969,Wolfe1971}
can be used to adjust the step size to avoid understepping or overstepping, and to maintain a 
positive definite $\bm{B}_k$.\cite{NocedalBook}

\subsection{Parallel transport for CSF orbital coefficients}

\begin{figure}[b]
	\includegraphics[width=\linewidth]{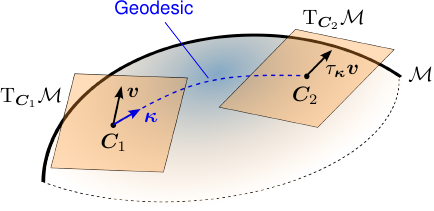}
	\caption{On a curved Riemannian manifold $\mathcal{M}$, the tangent space $\mathcal{T}_{\bm{C}}\mathcal{M}$ 
		changes at each point $\bm{C}$.
		Tangent vectors $\bm{v} \in \mathcal{T}_{\bm{C}}\mathcal{M}$ must be parallel transported 
		along the geodesic defined by a step direction $\bm{\kappa}$ such that the 
		transported vector $\tau_{\bm{\kappa}} \bm{v}$ remains in the tangent space at the new point.}
	\label{fig:geom}
\end{figure}

Since the CSF orbital coefficients $\bm{C}$ are constrained to a curved flag manifold $\mathcal{M}$, the 
gradient and orbital rotation step must lie in the linear tangent space $\mathrm{T}_{\bm{C}} \mathcal{M}$ at any given point. 
This geometric constraint is illustrated in Fig.~\ref{fig:geom}.
Updating the orbital coefficients corresponds to moving along the geodesic in  the
direction defined by the tangent step direction $\bm{\kappa}$.
The curvature of the flag manifold means that the tangent space changes at every point $\bm{C}$, 
and thus the gradients $\bm{g}_k$ and steps $\bm{s}_k$ from previous L-BFGS iterations must be parallel transported
into the tangent space for the current position $\bm{C}_k$ before $\bm{B}_{k}$ is evaluated (see Fig.~\ref{fig:geom}).
The formula to parallel transport a tangent vector $\bm{v} \in \mathrm{T}_{\bm{C}} \mathcal{M}$ 
on a flag manifold along the geodesic defined by $\bm{\kappa}$ has
previously been derived for the high-spin ROHF case by Vidal \textit{et al},\cite{Vidal2024} and can be directly extended to an arbitrary CSF. 
Specifically, the parallel transported vector $\tau_{\bm{\kappa}} \bm{v}$ can be 
computed as 
\begin{equation}
\tau_{\bm{\kappa}} \bm{v} =  \sum_{k=0}^{\infty} \frac{1}{k!} \qty(-\frac{1}{2})^k 
\qty[\bm{\kappa}, \qty[\cdots , \qty[\bm{\kappa}, \bm{v}]]]^k_\mathrm{T},
\label{eq:parallel_transport}
\end{equation}
where $\qty[\bm{\kappa}, \qty[\cdots , \qty[\bm{\kappa}, \bm{v}]]]^k_\mathrm{T}$ 
denotes a $k$-fold nested set of projected commutators.\cite{Vidal2024,Abe2013}
Each individual commutator $\qty[\bm{\kappa}, \bm{v}]_\mathrm{T}$ denotes a
standard matrix commutator $[\bm{\kappa}, \bm{v}]$ projected into the tangent space $\mathrm{T}$ such that
the components of $[\bm{\kappa}, \bm{v}]$  that connect two orbitals in the same shell are removed 
and $[\bm{\kappa}, \bm{v}]_\mathrm{T}$  adopts the block off-diagonal 
structure shown in Eq.~\eqref{eq:tangent_form}.
The cost to evaluate $\qty[\bm{\kappa}, \bm{v}]_\mathrm{T}$ scales as $\mathcal{O}(\Nb^3)$.
In practice, Eq.~\eqref{eq:parallel_transport} can be evaluated recursively and truncated 
when the maximum element in the highest-order nested projected commutator falls below a threshold value.
Since the orbital rotation step $\bm{\kappa}$ is generally small, this recursive scheme typically converges in at most 5 steps
with a threshold of $10^{-4}$.
Therefore, the cost of parallel transport is negligible compared to evaluating the Coulomb and exchange matrices.

If the CSF is a closed-shell determinant, then the orbital coefficients are constrained to a Grassmann manifold,\cite{Voorhis2002} 
which is a special case of a flag manifold with only two invariant subspaces (the occupied and virtual shells).
In that case, Eq.~\eqref{eq:parallel_transport} leaves tangent vectors $\bm{v}$ unchanged 
and the inverse Hessian update can be performed 
directly using $\bm{s}_k$ and $\bm{g}_k$ from previous iterations without parallel transport.\cite{Voorhis2002}
However, this property is no longer satisfied for an open-shell CSF, 
and thus the previous steps and gradients required for the L-BFGS 
update should be parallel transported on every iteration.

\subsection{Preconditioner and energy-weighted coordinates}
\label{sec:preconditioner}
The convergence of quasi-Newton algorithms can be significantly accelerated by using a preconditioner 
to ensure that the approximate inverse Hessian closely approximates the identify matrix.
For the CSF-GDM optimization, two methods are combined to achieve successful preconditioning. 
Firstly, the initial inverse Hessian is defined as a scaled identity matrix\cite{NocedalBook}
\begin{equation}
\bm{B}_{k+1}^0 = \gamma_{k+1} \bm{I},
\label{eq:H0gamma}
\end{equation}
where $\gamma_k$ approximates the local curvature as
\begin{equation}
\gamma_{k+1} = \frac{\bm{s}_{k}^\dagger \bm{y}_{k}^{\vphantom{\dagger}}}{\bm{y}_{k}^\dagger \bm{y}_{k}^{\vphantom{\dagger}}}.
\end{equation}
Secondly, a transformation into pseudo-canonical energy-weighted coordinates (EWCs) is used to construct 
a local coordinate system for the tangent space that approximately diagonalizes the Hessian.\cite{HeadGordon1988,Voorhis2002}

Constructing pseudo-canonical EWCs requires an approximation to the diagonal 
elements of the Hessian $Q_{pq,pq}$.
From the analytic expression [Eq.~\eqref{eq:Qpqrs}], the diagonal terms are given exactly as\cite{Chaban1997}
\begin{equation}
\begin{split}
Q_{pq,pq} &= 2(F^\cP_{qq} - F^\cQ_{qq}) + 2(F^\cQ_{pp} - F^\cP_{pp}) 
\\
&\quad+4(a_{pp} + a_{qq} - 4a_{pq})\expval{pq|qp}
\\
&\quad+2(b_{pp} + b_{qq} - 2a_{pq})\qty(\expval{pq|qp} + \expval{pq|pq}).
\end{split}
\label{eq:Qpqpq1}
\end{equation}
Following, Ref.~\onlinecite{Chaban1997}, the explicit two-electron integrals must be included for $Q_{wq,wq}$
when $w$ is an open-shell orbital, since the energy change for transformations involving open-shell orbitals can be small.
The relevant Coulomb and exchange integrals $\expval{wq|wq}$ and $\expval{wq|qw}$ can be 
evaluated using a $J$ and $K$ matrix build for each open-shell orbital with overall scaling 
$\mathcal{O}(\No \Nb^4)$.
These two-electron integrals can be ignored for core--virtual rotations, giving
\begin{equation}
Q_{ia,ia} \approx 2(F^\cC_{aa} - F^\cA_{aa}) + 2(F^\cA_{ii} - F^\cC_{ii}),
\label{eq:Qpqpq2}
\end{equation}
where $\cC$ ($\cA$) denotes the core (virtual) orbital shell.

Two transformations can now be performed to bring the Hessian into an approximate diagonal form. 
The first is to construct pseudo-canonical orbitals by transforming the orbitals within each invariant subspace (orbital shell) to 
diagonalize the generalized Fock matrix within that subspace, i.e., to bring $F^{\cP}_{pq}$ into a diagonal form for $p,q \in \cP$.
This canonical transformation can be achieved using an orthogonal transformation $\bm{C} \rightarrow \bm{C} \bm{U}$ 
with block diagonal form, e.g., for a CSF with four shells
\begin{equation}
\bm{U} = 
\begin{pmatrix}
\bm{U}_{1} & \bm{0}     & \bm{0}     & \bm{0} \\
\bm{0}     & \bm{U}_{2} & \bm{0}     & \bm{0} \\
\bm{0}     & \bm{0}     & \bm{U}_{3} & \bm{0} \\
\bm{0}     & \bm{0}     & \bm{0}     & \bm{U}_{4}
\end{pmatrix}.
\end{equation}
Since the generalized Fock matrix is zero for orbitals in the virtual  shell, the corresponding 
invariant transformation must be approximated by diagonalising the standard Fock matrix in the virtual--virtual subspace
\begin{equation}
f_{ab} = h_{ab}  + \sum_{r} n_r ( \expval{ar|br} - \frac{1}{2}\expval{ar|rb})
\end{equation}
In the second transformation, the tangent space coordinates are rescaled with a preconditioner \begin{equation}
\alpha_{pq} = \max(\abs{Q_{pq,pq}}^{\frac{1}{2}}, t_\alpha)
\label{eq:precdef}
\end{equation} 
built from
$Q_{pq,pq}$ in the pseudo-canonical orbital basis [Eqs.~\eqref{eq:Qpqpq1} and \eqref{eq:Qpqpq2}]
such that the approximate inverse Hessian becomes close to the identity matrix.
These transformations can be combined by transforming the gradient and orbital rotation steps from previous 
iterations as 
\begin{equation}
\bm{s}_k \rightarrow \bm{U}^\dagger \bm{s}_k  \bm{U}
\quad\text{and}\quad
\bm{g}_k \rightarrow \bm{U}^\dagger \bm{g}_k  \bm{U},
\label{eq:ewc_update}
\end{equation}
and then rescaling the step and gradient components as
\begin{equation}
\tilde{s}_{pq} = \frac{{s}_{pq}}{ \alpha_{pq}}
 \quad\text{and}\quad
\tilde{g}_{pq} =  \alpha_{pq}\, {g}_{pq}.
\label{eq:ewc}
\end{equation}
In Eq.~\eqref{eq:precdef}, the use of positive definite $\alpha_{pq}$ enforces downhill step directions,
while the threshold $t_\alpha $ (set here as $0.1$)  prevents step components from blowing up in Eq.~\eqref{eq:ewc}.
Once the quasi-Newton step has been evaluated,
the new orbital rotation step $\bm{s}_{k+1}$ in the pseudo-canonical basis 
is computed from the energy-weighted coordinates as
\begin{equation}
s_{pq} = \alpha_{pq}\, \tilde{s}_{pq}.
\label{eq:rescale}
\end{equation}
Preliminary tests showed that transforming to EWCs through 
Eq.~\eqref{eq:ewc} is essential to achieve satisfactory convergence and should be performed on every iteration,
whereas pseudo-canonicalization using Eq.~\eqref{eq:ewc_update} can be performed intermittently.

\subsection{Choice of initial orbitals}
\label{sec:initial_orbitals}

CSF-based ROHF theory can strongly depend on the choice of initial orbitals for certain spin coupling patterns, 
as shown by the numerical results  in Section~\ref{sec:results} (\textit{vide infra}) and  in Ref.~\citenum{Gouveia2024}.
In this work, two strategies are considered to identify the initial orbital coefficients. 
The first is to initialize the coefficients by performing random orbital rotations from the optimized high-spin ROHF solution, 
allowing multiple CSF solutions to be searched.
The second is to localize the high-spin ROHF orbitals  using the Pipek--Mezey method\cite{Pipek1989} (or another localization scheme)
and then assign these to the open shells such that the CSF energy is minimized (Appendix~\ref{sec:optimal_ordering}).
Alternative starting guess orbitals described in Ref.~\citenum{Gouveia2024} make use of the
atomic valence active space\cite{Sayfutyarova2017} procedure and spin-averaged HF,\cite{Stavrev1998} but these are not considered here.


\subsection{Outline of the CSF-GDM algorithm}
\label{sec:CSF_GDM}
Bringing together these components now allows the 
CSF-GDM algorithm to be defined as follows.
Starting from some suitable initial orbital coefficients, each iteration proceeds as:
\begin{enumerate}[itemsep=0em]
\item{Pseudo-canonicalize MOs and update previous steps $\{\bm{s}_k\}$ and gradients $\{\bm{g}_k\}$ 
using Eq.~\eqref{eq:ewc_update};}\label{it:ewc}
\item{Compute gradient $\bm{g}_{k+1}$ using the pseudo-canonical orbitals;}
\item{Convert all $\{\bm{s}_k\}$ and  $\{\bm{g}_k\}$ to energy-weighted coordinates using Eq.~\eqref{eq:ewc};}
\item{Compute the initial inverse Hessian $\bm{B}^0_{k+1}$ using Eq.~\eqref{eq:H0gamma};}
\item{Compute the quasi-Newton step $\tilde{\bm{s}}_{k+1}$ using the standard L-BFGS update formula;}
\item{Convert $\tilde{\bm{s}}_{k+1}$ to an orbital rotation using Eq.~\eqref{eq:rescale};}
\item{Update the MO coefficients using Eq.~\eqref{eq:MOupdate} with $\bm{\kappa} = \bm{s}_{k+1}$ 
and compute the new energy $E_{k+1}$;} \label{it:update}
\item{Parallel transport previous steps $\{\bm{s}_k\}$ and gradients $\{\bm{g}_k\}$ to the new origin using  Eq.~\eqref{eq:parallel_transport};}
\item{If not converged, return to step~\ref{it:ewc}.}
\end{enumerate}
The overall computational scaling is given by $\mathcal{O}(\No n^4)$, 
which is determined by the cost of computing
the Coulomb and exchange integrals in the preconditioner Eq.~\eqref{eq:Qpqpq1}.
All required integrals can be computed using standard routines to compute Coulomb and exchange matrices,
meaning that the algorithm can benefit from existing highly-optimized Fock-build routines.


\section{Computational details}
\label{sec:comp}
A pilot version of the CSF-GDM algorithm has been implemented in the \textsc{Quantel} library,\cite{Quantel} 
a Python/C++  package for developing electronic structure algorithms.
An interface to the \textsc{PySCF} package\cite{PySCFb} 
is used to evaluate all AO integrals and to construct the necessary Coulomb and exchange matrices.
Optimization steps are rescaled to satisfy the criterion $\norm{\bm{s}}_\infty \le 0.5$  
and the maximum number of previous steps in the  L-BFGS update is 20.
CSF-GDM calculations were converged to the criteria $\norm{\bm{g}}_\infty \le 10^{-6}\,\mathrm{E_h}$.
Benchmark CSF-ROHF calculations were performed in ORCA~6.0 using the comparable \texttt{VeryTightSCF} convergence 
criteria.\cite{Neese2022} 
The maximum number of iterations for all calculations was 1000.


\section{Results}
\label{sec:results}


\subsection{Convergence performance for transition-metal compounds}
\subsubsection{Mono-nuclear hexa-aquo complexes}
To assess the performance of the CSF-GDM algorithm, 
optimization statistics are compared against the recently introduced CSF-ROHF approach.
\cite{Gouveia2024}
The test set includes the different open-shell spin coupling states for the 3d transition 
metal hexa-aquo complexes 
with varying oxidation states, as listed in Table~\ref{tab:aqua_complexes}, using the def2-SVP basis set.\cite{Weigend2005}
The geometry of each complex is optimized for the high-spin state in the oxidation state for which the $\mathrm{t_{2g}}$ or
$\mathrm{e_g}$ crystal field orbitals have equal occupation to prevent any Jahn--Teller distortion.
Details of geometry optimisation, and the corresponding structures, are provided in \suppI{} Section~S1.
The same structure is then used for all oxidation states with the same metal centre.
The importance of parallel transport and pseudo-canonicalization is considered by performing CSF-GDM 
without these features, labelled as ``no PT'' and ``no PC'' respectively.
If no parallel transport or pseudo-canonicalization are included, 
then the CSF-GDM approach reduces to standard L-BFGS. 

\begin{table}[htb]
\caption{Spin coupling vectors considered for the 3d transition metal hexa-aquo complexes. 
34 different spin coupling vectors and complexes are considered in total.}
\label{tab:aqua_complexes}
\begin{ruledtabular}
\begin{tabular}{lcccc}
Complex & Singlet & Doublet & Triplet & Quartet \\
\hline
\multirow{2}{*}{\ce{[V(H2O)6]^2+}}   &       & $++-$ & & \\
                                    &       & $+-+$ & & \\ 
\hline
\ce{[V(H2O)6]^3+}   & $+-$  & & & \\ 
\hline
\multirow{3}{*}{\ce{[Cr(H2O)6]^2+}} &  $++--$ &            & $+++-$ &   \\
                                   &  $+-+-$ &             & $++-+$ &  \\
                                   &                &            & $+-++$ &  \\
\hline
\multirow{2}{*}{\ce{[Cr(H2O)6]^3+}}   &       & $++-$ & & \\
                                    &       & $+-+$ & & \\ 
\hline
\multirow{5}{*}{\ce{[Mn(H2O)6]^2+}} &                 & $+++--$ &                 & $++++-$ \\                            
                                     &                 & $++-+-$ &                 & $+++-+$ \\                   
                                     &                & $+-++-$ &                 & $++-++$ \\         
                                     &                & $++--+$ &                 & $+-+++$ \\         
                                     &                & $+-+-+$ &                 &                  \\        
\hline
\multirow{3}{*}{\ce{[Fe(H2O)6]^2+}} &  $++--$ &            & $+++-$ &  \\
                                   &  $+-+-$ &             & $++-+$ &  \\
                                   &                &            & $+-++$ &  \\
\hline
\multirow{5}{*}{\ce{[Fe(H2O)6]^3+}} &                 & $+++--$ &                 & $++++-$ \\                            
                                     &                 & $++-+-$ &                 & $+++-+$ \\                   
                                     &                & $+-++-$ &                 & $++-++$ \\         
                                     &                & $++--+$ &                 & $+-+++$ \\         
                                     &                & $+-+-+$ &                 &                  \\    
\hline
\ce{[Ni(H2O)6]^2+}   & $+-$  & & & \\ 
\end{tabular}
\end{ruledtabular}
\end{table}

Statistics for the mean, median, minimum, and maximum number of iterations  required to reach convergence starting from the high-spin ROHF orbitals 
are presented in Table~\ref{tab:aqua_convergence}.
\begin{table}[b]
\caption{Optimization statistics for the 3d transition-metal hexa-aquo complexes (see Table~\ref{tab:aqua_complexes}) 
starting from the high-spin ROHF orbitals.
\label{tab:aqua_convergence}}
\begin{ruledtabular}
\begin{tabular}{lcccccc}
\multirow{2}{*}{Algorithm}  & \multirow{2}{*}{Mean} & \multirow{2}{*}{Median} & \multirow{2}{*}{Min} & \multirow{2}{*}{Max} & Local  & \multirow{2}{*}{Fail} \\
                  &  &  &  &  &  Minima  & \\
\hline
CSF-GDM                     & 42.4 & 45.0 & 9  & 104 & 0 & 0 \\
CSF-GDM (no PC)       & 43.9 & 47.0 & 9  & 104 & 1  & 0 \\
CSF-GDM (no PT)       & 44.1 & 46.0 & 9  & 102 & 1 & 0 \\
L-BFGS                        & 44.8 & 49.0 & 9  & 101 & 0 & 0 \\ 
\hline
CSF-ROHF                   & 80.6 & 45.0 & 13 & 940 & 23 & 2\\ 
\end{tabular}
\end{ruledtabular}
\end{table}
All the CSF-GDM variants converge in fewer iterations than the CSF-ROHF algorithm, while the latter fails to converge to the lowest
energy solution in 23 out of 34 cases.
The optimization is deemed to have converged to a local minimum if it does not find the lowest energy
solution found across all algorithms, 
with a threshold of $1\,\mathrm{\upmu  E_h}$ used to identify equivalent local minima.
CSF-GDM generally converges in slightly fewer iterations when both pseudo-canonicalization and parallel transport are included, 
although this effect is relatively small.
Converged CSF energies are tabulated in \suppI{} Table~S1.

It is surprising that pseudo-canonicalization and parallel transport only slightly improve
the convergence behaviour, even though they more accurately account for the curvature of the ROHF manifold.
One possibility is that these algorithmic components are more significant when the optimization starts further away from convergence. 
To test this hypothesis, convergence statistics were also obtained  using a core orbital guess, which is expected to be a worse 
starting point than the high-spin ROHF orbitals (Table~\ref{tab:aqua_convergence_core}).
Converged CSF energies are tabulated in \suppI{} Table~S2.
Including both pseudo-canonicalization and parallel transport now reduces the mean number of iterations  
by 7\,\% compared to standard L-BFGS, and leads to fewer local minima.  
Furthermore,  CSF-ROHF  becomes much less robust for these less accurate starting guesses, 
failing to converge nine times, and leading to more iterations and local minima than the CSF-GDM approach. 
These results demonstrate the advantage of quasi-Newton optimization methods for 
CSF-based ROHF calculations, and the importance of a good initial guess.

\begin{table}[htb]
	\caption{Optimization statistics for the 3d transition-metal hexa-aquo complexes (see Table~\ref{tab:aqua_complexes}) 
		starting from a core orbital guess.}
	\label{tab:aqua_convergence_core}
	\begin{ruledtabular}
		\begin{tabular}{lcccccc}
			\multirow{2}{*}{Algorithm}  & \multirow{2}{*}{Mean} & \multirow{2}{*}{Median} & \multirow{2}{*}{Min} & \multirow{2}{*}{Max} & Local  & \multirow{2}{*}{Fail} \\
			&  &  &  &  &  Minima &  \\
			\hline
			CSF-GDM                     & 119.6 & 118.5 & 90 & 167 & 10 & 0 \\
			CSF-GDM (no PC)       & 132.4 & 127.0 & 89 & 221 & 12 & 0 \\
			CSF-GDM (no PT)       & 123.4 & 120.0 & 93 & 163 & 12 & 0 \\
			L-BFGS                        & 128.5 & 131.0 & 86 & 170 & 18 & 0 \\ 
			\hline
			CSF-ROHF      	          & 144.3 & 100.0 & 22 & 437 & 23 & 9 \\ 
		\end{tabular}
	\end{ruledtabular}
\end{table}

\subsubsection{Benchmark \ce{Cr2} and \ce{CrC} molecules}

Next, the convergence performance of CSF-GDM is tested for the \ce{Cr2} and \ce{CrC} diatomic molecules,
 which have become challenging benchmarks for new SCF optimisation
algorithms.\cite{Slattery2024,Dittmer2023}
Following Ref.~\citenum{Slattery2024}, the def2-TZVPP basis set\cite{Weigend2005} is used with a bond length of $\SI{2}{\angstrom}$ for both molecules.
The number of iterations ($N_\text{Iter}$)  required to converge various high- and low-spin configurations
with different initial guesses is reported in Table~\ref{tab:Cr2}.

For \ce{Cr2}, the convergence of the closed-shell RHF state using CSF-GDM in 203 iterations is competitive with other recent second-order
SCF algorithms, such as the Quasi-Newton Unitary Optimization with Trust-Region (QUOTR) algorithm, which converges to the same solution 
in 193 iterations.\cite{Slattery2024}
In contrast, the CSF-ROHF algorithm in ORCA~6.0, which uses standard DIIS optimization, failed to locate an RHF solution, although more sophisticated algorithms, such as the 
trust-radius augmented Hessian method,\cite{HelmichParis2021} were successful.
For each high-spin ROHF configuration, CSF-GDM finds a lower energy solution than CSF-ROHF. 
The advantage of using localised high-spin ROHF initial orbitals that are ordered to minimise the 
CSF energy is shown by the rapid convergence of the antiferromagnetic [+++++\textminus\textminus\textminus\textminus\textminus] CSF in 12 iterations,
compared to 43 iterations using the canonical high-spin ROHF orbitals (the core initial guess
converged to a local minimum).
By comparison, CSF-ROHF failed to converge for this antiferromagnetic  CSF
starting from the core orbital guess.

\begin{table*}[htb]
\caption{Comparison of CSF-GDM and CSF-ROHF for \ce{Cr2} and \ce{CrC} at $\SI{2.0}{\angstrom}$ bond length with selected CSFs, using the core or high-spin ROHF initial guess. When a calculation failed to converge in 1000 iterations, 
the energy of the last iteration is reported in italics.}
\label{tab:Cr2}
	\begin{ruledtabular}
		\begin{tabular}{lrccccc}
	   & \multirow{2}{*}{Spin Coupling} & \multirow{2}{*}{Initial Guess} & \multicolumn{2}{c}{CSF-GDM} & \multicolumn{2}{c}{CSF-ROHF (ORCA 6.0\cite{Neese2022})} \\
	   \cline{4-5} \cline{6-7}
       & &  & $N_\text{Iter}$ &  $E_{\text{min}} / \mathrm{E_h}$ & $N_\text{Iter}$ &  $E_{\text{min}} / \mathrm{E_h}$ \\
      \hline 
    \multirow{9}{*}{\ce{Cr2}}  & RHF  & core & 209  & $-2086.159\,612$ & \textit{fail} & $\mathit{-2085.872\,496}$  \\
    &  $[++]$          & core & 198 & $-2086.213\,166$ & 29 & $-2085.876\,204$  \\
    &  $[++++]$        & core & 206 & $-2086.260\,779$ & 73 & $-2085.921\,253$  \\
    &  $[++++++]$      & core & 169 & $-2086.309\,706$ & 38 & $-2085.882\,240$  \\
    &  $[++++++++]$    & core & 130 & $-2086.426\,127$ & 617 &  $-2086.075\,782$ \\
    &  $[++++++++++]$  & core & 75  & $-2086.527\,987$ & 30 & $-2086.366\,815$ \\ 
    &  $[+++++-----]$  & core & 127 & $-2086.177\,303$ & \textit{fail} & $\mathit{-2085.917\,024}$  \\
    &  $[+++++-----]$  & ROHF & 43  & $-2086.506\,284$ & -- & --   \\
    &  $[+++++-----]$  & ordered ROHF  & 12 & $-2086.506\,284$ & -- & -- \\
\hline
    \multirow{9}{*}{\ce{CrC}}   & RHF & core & 36 & $-1080.566\,167$ & 52 & $-1080.698\,666$\\
      & $[++]$       & core & 165  & $-1080.822\,232$ & 70 & $-1080.787\,931$ \\
      & $[++++]$     & core & 139 & $-1080.880\,312$  & 269 & $-1080.902\,882$ \\
      & $[++++++]$   & core & 131 & $-1080.967\,996$  & 136 & $-1080.967\,996$ \\
      & $[++++++++]$ & core & 98 & $-1081.061\,727$   & 71 & $-1081.061\,727$ \\
      & $[++++----]$ & core & 115 & $ -1080.855\,416$ & 234 & $-1080.819\,732$\\
      & $[++++----]$ & ROHF & 26 & $-1081.005\,627$   & -- & -- \\
      & $[++++----]$ & ordered ROHF  & 12 & $-1081.005\,627$ & -- & -- \\
	\end{tabular}
	\end{ruledtabular}
\end{table*}

Similar performance is observed for the \ce{CrC} molecule,
where now the low-spin ROHF convergence is demonstrated with the   [++++\textminus\textminus\textminus\textminus] spin coupling.
CSF-GDM converges in only 36 iterations for the closed-shell RHF state, but finds a higher energy 
local minimum compared to Ref.~\citenum{Slattery2024} 
and the solution obtained from ORCA~6.0.
Again, using localized and optimally ordered  high-spin ROHF orbitals as the initial guess
leads to a lower energy antiferromagnetic CSF compared to the core orbital guess, and converges in half as many iterations compared to starting from the canonical high-spin ROHF  orbitals.
In contrast, CSF-ROHF requires 234 iterations to converge for the antiferromagnetic CSF
state using the core orbital guess, and finds a higher-energy solution.

\subsection{Multiple solutions in iron-sulfur complexes}

It is well-known that single determinant methods exhibit local minima,
particularly when a single determinant is used to describe a multiconfigurational state. 
Since there is only a small change in CSF energy when the assignment 
of open-shell orbitals to different shells is varied, 
it is likely that more local minima occur when there are many singly-occupied shells. 
The ability to systematically converge minima from various starting points using CSF-GDM provides
an opportunity to ask: what is the nature and frequency of local minima in low-spin ROHF theory?
This question is investigated using 
the synthetic iron-sulfur complexes  \ce{[Fe(SCH3)_4]^-} and  \ce{[Fe_2S_2(SCH3)_4]^2-},
which provide models for iron-sulfur clusters
involved in biological redox processes and electron transfer.\cite{Mayerle1975,Sharma2014} 
Previous CSF energies for  \ce{[Fe(SCH3)_4]^-} with various spin coupling patterns 
are available from the literature for verification.\cite{Gouveia2024}
Geometries for both clusters were the same as Ref.~\onlinecite{Gouveia2024} (originally 
from Refs.~\onlinecite{Chilkuri2017} and \onlinecite{Chilkuri2019})
and the def2-TZVP basis set\cite{Weigend2005} was used.

\subsubsection{Mononuclear \ce{[Fe(SCH3)_4]^-}}
\begin{figure*}[htb]
\includegraphics[width=\linewidth]{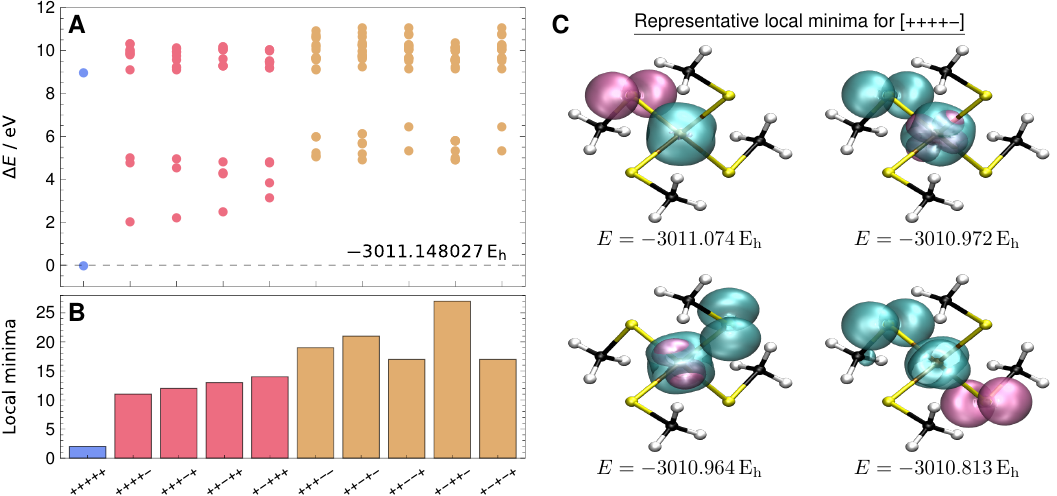}
\caption{(\textsf{\textbf{A}}) Energies of the lowest local minima found through random searches for each spin-coupling pattern in the \ce{[Fe(SCH3)_4]^-} complex (def2-TZVP), shown relative to the $S=5/2$ ROHF global minimum. Energies are tabulated in \suppI{} Table~S3.
(\textsf{\textbf{B}}) Local minima generally become more prevalent as the number of unique open shells increases.
(\textsf{\textbf{C}}) For the [++++\textminus] global minimum ($E=-3011.074\,\mathrm{E_h}$), 
the [\textminus] shell density (purple) is localized on a 
\ce{S} atom while the [++++] shell density (cyan) is localized on the \ce{Fe} centre (top left). 
Higher-energy local minima correspond the different arrangements of electrons localized on one or two \ce{S} atoms.
}
\label{fig:FeS}
\end{figure*}

For the mononuclear  \ce{[Fe(SCH3)_4]^-} complex, 100 independent CSF-GDM calculations were performed 
starting from different initial orbital coefficients.
These guesses were obtained by first converging the corresponding high-spin ROHF orbital coefficients
and then applying a random orbital rotation. 
This process was repeated for all spin coupling patterns with five unpaired electrons and 
the resulting local minima energies are shown in Fig.~\ref{fig:FeS}\textcolor{blue}{A}.
With only 100 initial guesses, these data are unlikely to provide an exhaustive set of solutions, 
but they are sufficient to illustrate the key properties of different local minima.
Degenerate solutions are counted as one minimum in the following analysis.

Only two minima are found for the high-spin ROHF calculation, with the lowest energy solution matching 
the result from Ref.~\onlinecite{Gouveia2024}.
The electronic structure of the global minimum corresponds to localizing all five unpaired electrons on the \ce{Fe} atom.
In contrast, for the higher-energy local minimum, one unpaired electron is localized on a \ce{S} atom.

For the spin-coupling patterns with $S=3/2$, the local minima roughly form three energy groupings.
The corresponding open-shell  structure can be visualized by plotting the electron density for 
each shell of unpaired electrons, as illustrated by the  density plots for the [++++\textminus] spin coupling in Fig.~\ref{fig:FeS}\textcolor{blue}{C}.
In the lowest energy solution ($E=-3011.074\,\mathrm{E_h}$), the [\textminus] shell (purple) is 
localized on a \ce{S} atom, with the remaining unpaired electrons localized on the \ce{Fe} centre.
This arrangement minimizes the repulsive exchange interaction 
between electrons in the [++++] and [\textminus] shells.
The next highest pair of solutions ($E=-3010.972\,\mathrm{E_h}$ and $E=-3010.964\,\mathrm{E_h}$) correspond to 
configurations where the electron in the [\textminus] shell is localized on the \ce{Fe}, 
with one of the [++++] electrons localized on a \ce{S} atom and the remainder on \ce{Fe}.
Finally, in the highest-energy group of local minima, 
there are two unpaired electrons localized on different \ce{S} atoms, 
as illustrated by the solution with energy $E=-3010.813\,\mathrm{E_h}$ in Fig.~\ref{fig:FeS}\textcolor{blue}{C}.
Similar patterns of electron localization are observed for the other $S=3/2$ spin-coupling patterns, 
and also for the CSFs with $S=1/2$.

The presence of several local minima with unpaired electrons localized on \ce{S} atoms hints
at important electron correlation processes in this complex. 
It is clear from these data, and previous results,\cite{Gouveia2024} that the non-Hund ordering of 
spin states in \ce{[Fe(SCH3)_4]^-} is incorrectly described at the ROHF level of theory.
The dominant correlation mechanisms that stabilize the low-spin states below the high-spin 
states are known to involve metal-to-ligand charge transfer, 
leading to partial open-shell character on the \ce{S} atoms.\cite{Malrieu2013}
This process manifests in CSF-based ROHF theory as low-energy minima with an unpaired electron in a \ce{S}
3p orbital, as shown in Fig.~\ref{fig:FeS}\textcolor{blue}{C}.
The existence of local CSF minima that mimic important correlation mechanisms is analogous to 
symmetry-broken unrestricted HF solutions that are found for strongly-correlated 
open-shell molecules,\cite{Trail2003,Burton2021a,Burton2021b} and has also been observed for
multiple CASSCF solutions.\cite{Marie2023,Saade2024}

These results strongly indicate that CSF-based ROHF calculations using complex spin-coupling patterns
are highly susceptible to local minima. 
Therefore, the success of this approach relies on selecting a good initial guess that incorporates the 
expected electron localization.
However, even if a good initial guess can be found, 
there is no guarantee that the expected electron configuration will form a local minimum of 
the CSF energy.
For example, despite initialising the open-shell orbitals using localized \ce{Fe} 3d orbitals obtained from a 
high-spin ROHF calculation, no local minimum was found with all the unpaired electrons 
localized on the \ce{Fe} atom for any of the $S=3/2$ or $S=1/2$ spin-coupling patterns.
This result suggests that the $S=3/2$ or $S=1/2$ solutions identified using CSF-ROHF in 
Ref.~\onlinecite{Gouveia2024} are saddle points of the CSF energy, which can be verified using
stability analysis.

\subsubsection{Bimetallic \ce{[Fe_2S_2(SCH3)_4]^2-}}
\begin{figure*}[htb]
\centering
\includegraphics[width=0.925\linewidth]{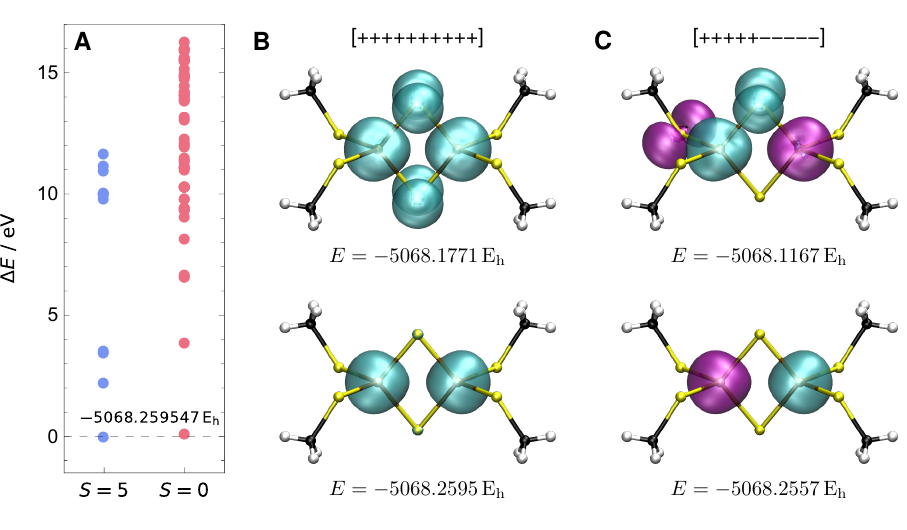}
\caption{
(\textsf{\textbf{A}}) Energies of the lowest local minima for the $S=5$ ferromagnetic  [++++++++++] and 
$S=0$ antiferromagnetic [+++++\textminus\textminus\textminus\textminus\textminus] spin-coupling 
in \ce{[Fe_2S_2(SCH3)_4]^2-} (def2-TZVP)
relative to the $S=5$ global minimum. Converged energies are tabulated in  \suppI{} Table~S4.
(\textsf{\textbf{B}}) Singly-occupied shell density plots for the two lowest energy $S=5$ minima 
show that the global minimum (bottom) only has unpaired electrons on \ce{Fe} atoms 
while the first local minimum (top) has open-shell character on \ce{S} atoms.
(\textsf{\textbf{C}}) Singly-occupied shell density plots for the two lowest energy $S=0$ minima show that
the global minimum (bottom) has the [+++++] shell (cyan) localized on one \ce{Fe} 
and the [\textminus\textminus\textminus\textminus\textminus] shell (purple) on the other \ce{Fe}.
Again, the first local minimum (top) has unpaired electrons on \ce{S} atoms.}
\label{fig:Fe2S2}
\end{figure*}

Next, the \ce{[Fe_2S_2(SCH3)_4]^2-} complex is considered as an example of a bimetallic cluster with an 
antiferromagnetic ground state.\cite{Sharma2014}
This larger cluster provides an opportunity to study the number of solutions as the number of 
open-shell electrons increases.
The ferromagnetic high-spin $S=5$ state can approximated by a CSF with the 
spin-coupling pattern [++++++++++], while the antiferromagnetic low-spin 
$S=0$ state can be approximated by the spin-coupling pattern [+++++\textminus\textminus\textminus\textminus\textminus] 
with the [+++++] shell  localized on one \ce{Fe} centre and the [\textminus\textminus\textminus\textminus\textminus] shell on the other \ce{Fe}.
It was previously shown that these CSF approximations alone do not predict the correct 
energetic ordering of the two spin states.\cite{Gouveia2024}
Correctly predicting the antiferromagnetic ground state requires additional 
electron correlation mechanisms involving charge transfer and ionic configurations,\cite{Malrieu2013}
which are not present in the CSF approximation.

Following the same procedure as the mononuclear complex, 
100 independent CSF-GDM calculations were performed using different starting points obtained 
through a random perturbation to the lowest-energy high-spin ($S=5$)  orbital coefficients.
The resulting local minima are plotted for the two spin states in Fig.~\ref{fig:Fe2S2}\textcolor{blue}{A}, 
with energies shown relative to the lowest-energy high-spin solution ($E=-5068.259547\,\mathrm{E_h}$).
Notably, for the low-spin CSF, the expected antiferromagnetic \ce{Fe}--\ce{Fe} coupling was not 
found using any of the 100 random starting points.
This surprising result indicates that the CSF energy landscape can have many local minima with 
small catchment basins, emphasising the dependence on the initial guess.
Instead, a physically-motivated guess was defined by first converging the lowest-energy 
high-spin solution, then localising the open-shell orbitals, and finally assigning these local orbitals.
to each shell in an order that minimizes the exchange interaction, 
using the algorithms described in Appendix~\ref{sec:optimal_ordering}.
The resulting solution was found to be the lowest energy antiferromagnetic CSF with the two open shells 
centred on separate \ce{Fe} atoms  (Fig.~\ref{fig:Fe2S2}\textcolor{blue}{C}, bottom), and is
nearly degenerate with the $S=5$ global minimum (Fig.~\ref{fig:Fe2S2}\textcolor{blue}{B}, bottom).

The lowest-energy solution for both spin states corresponds to the ferromagnetic ($E=-5068.2595\,\mathrm{E_h}$)
and antiferromagnetic ($E=-5068.2557\,\mathrm{E_h}$) coupling of the two \ce{Fe} atoms in local d\textsuperscript{5} configurations ($S=5/2$), as illustrated by singly-occupied shell density plots
in  Figs.~\ref{fig:Fe2S2}\textcolor{blue}{B} and \ref{fig:Fe2S2}\textcolor{blue}{C}.
These are the only two minima that were found with the open-shell electrons localized on only the \ce{Fe} atoms, 
and the energy difference of $843.845\,\mathrm{cm^{-1}}$ matches previous results.\cite{Gouveia2024}
In direct analogy with the mononuclear complex, the higher-energy local minima for \ce{[Fe_2S_2(SCH3)_4]^2-} 
correspond to electronic structures where one (or more) of the unpaired electrons is 
localized on a \ce{S} atom. 
This result is illustrated using the singly-occupied shell density plots for the first non-global
minimum for each spin coupling in Figs.~\ref{fig:Fe2S2}\textcolor{blue}{B} and
\ref{fig:Fe2S2}\textcolor{blue}{C}.

\begin{figure*}[htb]
	\includegraphics[width=\linewidth]{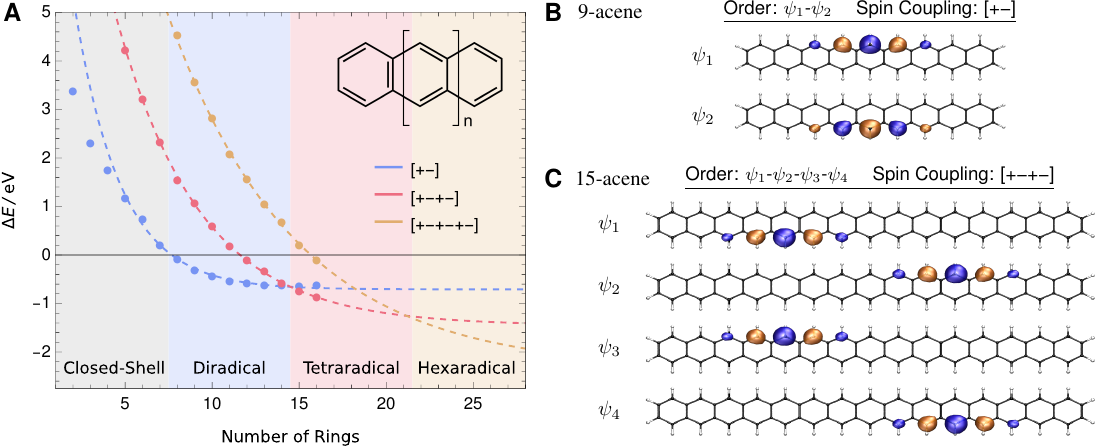}
	\caption{
		(\textsf{\textbf{A}}) Energies of the singlet ground state CSF solutions (cc-pVDZ) 
		with diradical [+\textminus], tetraradical [+\textminus+\textminus], and hexaradical [+\textminus+\textminus+\textminus]  
		spin coupling in polyacenes, plotted relative to the closed-shell RHF energy.
		Total energies are provided in \suppI{} Table~S5.
		Changes in the lowest energy CSF indicate the onset of polyradical character for larger polyacenes.
		(\textsf{\textbf{B}}) Open-shell orbitals for the diradical CSF [+\textminus] in 9-acene.
		(\textsf{\textbf{C}}) Open-shell orbitals for the tetraradical CSF [+\textminus] in 15-acene.
	} \label{fig:polyacene}
\end{figure*}

The increase in local minima and the larger spread of energies compared to the mononuclear complex 
can be explained by the larger number of \ce{S} atoms in \ce{[Fe_2S_2(SCH3)_4]^2-}. 
In particular, there are several ways to localize multiple unpaired electrons electrons onto
different \ce{S} environments, which leads to many non-degenerate CSF solutions. 
This result suggests that the number of local minima is likely to increase in larger complexes with more
diverse atomic environments. 
Finally, despite the small number of samples, it is surprising that the random search did not find the 
antiferromagnetic ground state solution.
These findings reinforce the importance of the initial guess in CSF optimization.

\subsection{Open-shell character of singlet ground state in polyacenes}
\label{sec:polyacenes}

Finally, the CSF approach can be used to qualitatively study the open-shell character 
in non-metallic compounds, such as organic molecules. 
This approach can be illustrated using the singlet ground state in polyacene chains (\ce{C_{4n+2}H_{2n+4}}), 
which are candidates for singlet fission materials\cite{Zimmerman2010,Zeng2014} 
or interstellar compounds,\cite{Hudgins1995}
and provide building blocks for nanographene fragments.\cite{Zeng2021,Pozo2024}
While high-accuracy wavefunction calculations,\cite{Ibeji2015,Schriber2016,Sharma2019,Meitei2021} 
including DMRG,\cite{Hachmann2007} have shown that polyacenes have a singlet ground state, 
the qualitative open-shell character in large polyacenes is not as easy to deduce from correlated calculations.
Spin-symmetry breaking in unrestricted KS-DFT has been proposed as evidence of a
diradical ground state in larger polyacenes,\cite{Bendikov2004} 
while a second symmetry breaking in polyacenes with 
13 or more rings indicates the onset of tetra-radical character.\cite{Trinquier2018}
The analysis of natural orbital occupation numbers from DMRG\cite{Hachmann2007} supports these findings,
but it has been argued that the presence of symmetry breaking in KS-DFT is not a reliable 
diagnostic for open shell character due to the possibility of artificial symmetry breaking,\cite{Hajgato2009} 
which is common in SCF calculations with significant HF exchange.\cite{Cohen2001,Pillai2025}

Optimising individual CSFs with different open-shell character provides an alternative route to probe
the qualitative nature of the singlet ground state without resorting to spin-symmetry breaking, while 
retaining mean-field computational cost.
In particular, the energies of CSF solutions with closed-shell, diradical, tetraradical, or hexaradical 
spin coupling can be compared to identify the most stable number of unpaired electrons in the ground state
and where they are localized in the molecule.
To test this idea, CSF-GDM calculations with various degrees of polyradical character were performed 
for the $n$-acene series up to $n=16$ using the cc-pVDZ basis set.\cite{Dunning1989} 
The largest calculation includes over 100 atoms and 1104 MOs, illustrating the scalability of CSF-GDM.
Molecular structures for $n=2$--6, 8, 10, and $12$ were taken from Ref.~\onlinecite{Hachmann2007},
while structures for $n=7$, 9, 11, and 13--16 were computed using UB3LYP/6-31G(d) in Q-Chem\cite{QChem54}
(see \suppI{} Section~S2).
Initial coefficients were obtained from the high-spin ROHF orbitals, which were localized
and assigned to shells using the algorithm outlined in Appendix~\ref{sec:optimal_ordering}.

The energies of the ground-state CSF solutions with diradical, tetraradical, or hexaradical 
spin coupling are shown relative to the corresponding RHF solution for each structure in Fig.~\ref{fig:polyacene}\textcolor{blue}{A}.
Only the [+\textminus+\textminus] tetraradical and  
[+\textminus+\textminus+\textminus] hexaradical solutions are shown, although the energetics 
are qualitatively unchanged if the other $S=0$ spin coupling patterns with 4 or 6 unpaired electrons are used 
instead (see \suppI{} Table~S5).
For the short polyacenes, the closed-shell solution is lowest in energy and the ground state is expected to be 
dominated by closed-shell character.
However, as the number of rings increases, the lowest-energy solution becomes the diradical 
CSF at 8-acene, and then becomes the tetradradical CSF for 15-acene. 
This transition in the lowest-energy CSF solution supports the onset of polyradical character 
for the 
longer polyacenes.

In Ref.~\onlinecite{Trinquier2018}, Trinquier and co-workers proposed that the longer acenes are composed of multiple
disjoint diradicals localized in tetra-methylene units linked by aromatic napthalene or benzene rings.
Their argument was motivated by the onset of KS-DFT spin symmetry breaking in 7-acene, with a second 
symmetry breaking at 13-acene.
Remarkably, this qualitative picture is corroborated by the relative energies of the different CSF states, which suggest a transition to diradical character around 7--8 rings, 
and to tetraradical character around 14-15 rings.
Extrapolation using an exponential fit  (dashed line Fig.~\ref{fig:polyacene}\textcolor{blue}{A}) 
suggests that the ground state becomes hexaradical at around 
21--22 rings, which again matches the prediction in Ref.~\onlinecite{Trinquier2018}.
Furthermore,  the open-shell orbitals for the diradical CSF in 9-acene (Fig.~\ref{fig:polyacene}\textcolor{blue}{B})
and the tetraradical CSF in 15-acene (Fig.~\ref{fig:polyacene}\textcolor{blue}{B}) support the predicted
localization of each unpaired electron to five carbons on one edge of the polyacene chain.
Notably, these local orbitals emerge directly from the CSF optimization 
without any post-SCF localization, in contrast to the approach described in Ref.~\onlinecite{Trinquier2018}.

These results suggest that CSF calculations can provide valuable insights into the localization of unpaired 
electrons, giving a conceptual understanding of open-shell electronic structures.
Assessing open-shell character using natural occupation numbers from correlated wavefunctions 
involves a degree of arbitrariness as the occupation number varies continuously from 0 to 2. 
On the other hand, unphysical ``artificial'' spin-symmetry breaking in KS-DFT can suggest open shell
character in molecules that would be considered closed-shell, such as benzene.\cite{Cohen2001,Lee2019,Pillai2025}
A single mean-field CSF provides an intermediate picture that retains spin symmetry, 
naturally results in localized unpaired electrons, and has fixed natural orbital occupations of 0, 1, or 2.
Potential applications of this low-cost methodology might include graphene 
nanoflakes beyond the reach of active space studies, 
where unpaired electrons lead to small energy gaps and exotic magnetic properties.\cite{Song2021,Pozo2024}




\section{Conclusion}
\label{sec:conclusions}
CSF-based ROHF theory is a promising approach to obtain compact reference states for open-shell problems 
without relying on CASSCF theory.
However, finding the optimal orbitals for a given CSF requires new optimization algorithms for 
an arbitrary number of open shells.
In this work, I have introduced a quasi-Newton Riemannian  optimization algorithm ``CSF-GDM''
that enables robust energy minimization for CSFs with arbitrary genealogical spin coupling.
This approach takes into account the structure of the CSF orbital constraint manifold and provides 
a generalized open-shell extension to the single-determinant GDM algorithm.\cite{Voorhis2002}
Compared to the CSF-ROHF approach introduced recently in Ref.~\citenum{Gouveia2024}, 
the CSF-GDM algorithm generally converges in fewer iterations, is much less likely to 
find higher-energy local minima, and avoids saddle points of the energy.

Robust energy minimization using CSF-GDM has allowed important properties of CSF-based calculations 
to be investigated.
Using the \ce{[Fe(SCH3)_4]^-} and \ce{[Fe_2S_2(SCH3)_4]^2-} complexes as illustrative examples, 
I have shown that there can be many local minima on 
the CSF energy landscape, particularly for a CSF with several distinct open shells.
These higher-energy solutions mimic key correlation processes, 
such as ligand-to-metal charge transfer, providing some qualitative insight into the 
true electronic structure of the ground state.
However, the presence of many local minima emphasizes the importance of finding a good initial 
orbital guess that reflects the physical open-shell character and long-range electron correlation.
More detailed studies into the electronic energy landscape of CSF-based ROHF theory will be required 
to fully characterise the properties and impact of these multiple solutions.

Furthermore, I have shown how mean-field CSF calculations can be used to gain qualitative insights 
into the localization of unpaired electrons in open-shell ground states.
Comparing the relative energies of CSF solutions with different numbers of unpaired electrons 
revealed that the singlet ground state of polyacene chains becomes progressively polyradical 
as the number of rings increases, in line with previous predictions.\cite{Trinquier2018}
The ability of CSF-based ROHF theory to provide spin-pure open-shell solutions, which automatically
localize unpaired electrons where appropriate, creates a valuable new tool for studying polyradical 
molecules and magnetic compounds.
Crucially, it does not rely on mean-field spin-symmetry breaking, which can be an 
unreliable indicator of open-shell character due to the possibility of 
artificial symmetry breaking.\cite{Cohen2001,Davidson1983,Lee2019,Pillai2025}

Moving forwards, the ability to optimize low-spin open-shell configurations with an arbitrary 
number of unpaired electrons at mean-field cost creates several opportunities to develop new 
methodology.
CSF states have already been successfully used  to create minimal multiconfigurational reference states and a sparse
Hilbert space for approximate CI calculations,\cite{Manni2020,Dobrautz2021,Dobrautz2022,Manni2021a,Manni2021b,Gouveia2024b}
and to define high-fidelity initial states for future quantum
algorithms.\cite{MartiDafcik2024,MartiDafcik2025,Sugisaki2016,Sugisaki2019} 
In addition, it would be interesting to consider new single-reference correlation theories
by applying many-body perturbation theory directly to an optimized CSF solution,
providing a route towards quantitative energies for low-energy spin states, 
or ionization energies and electron affinities in open-shell systems.\cite{Ammar2024}

\section*{Supplementary Material}
The \suppI{} file (PDF) includes
details of geometry optimization for the transition metal hexa-aquo complexes and polyacene 
chains, with optimized molecular coordinates,
and converged CSF energies for all systems considered.

\section*{Acknowledgements}
H.G.A.B was supported by Downing College, Cambridge through the Kim and Julianna Silverman Research Fellowship
and is a Royal Society University Research Fellow (URF\textbackslash{}R1\textbackslash{}241299) at University College London.

\section*{Data Availability}
The data that supports the findings of this study are available within the article and its
supplementary material.
All numerical data and molecular structures are available in a publicly available repository  [\href{https://doi.org/10.5281/zenodo.16738382}{10.5281/zenodo.16738382}].

\appendix

\section{Derivation of analytic Hessian terms}
\label{sec:hessian}
Following Ref.~\citenum{HelgakerBook}, the analytic second derivatives of the energy for an arbitrary wave function are given by
\begin{equation}
Q_{pqrs} 
= P_{pq}P_{rs}[2\gamma_{pr} h_{qs} - \delta_{qs}(F^{\cP}_{pr} + F^{\cR}_{rp}) + 2 Y_{pqrs} ],
\label{eq:hessian}
\end{equation}
where $P_{pq} = 1 - (pq)$ introduces an antisymmetric permutation of the indices $p$ and $q$, 
and $F_{pq}$ are the elements of the generalized Fock matrix defined in Eq.~\eqref{eq:genfock_JK}.
The intermediate terms $Y_{pqrs}$ are defined as
\begin{equation}
Y_{pqrs} = \sum_{mn} \qty[ \qty(\Gamma_{prmn} + \Gamma_{pnmr}) \expval{qn|ms} + \Gamma_{pmrn} \expval{qm|sn}].
\label{eq:Yint}
\end{equation}
Explicit expressions for the $Y_{pqrs}$ terms can be obtained using the structure of the CSF density matrices.
For a single CSF, the only non-zero 2-RDM terms are $\Gamma_{pqpq}$ and $\Gamma_{pqqp}$, leading to the expression
\begin{equation}
\begin{split}
Y_{pqrs} &=\delta_{pr}\sum_{m}\qty(\Gamma_{pmpm}\expval{qm|sn} + \Gamma_{pmmp}\expval{qm|ms}) \\
&+\delta_{pr} \Gamma_{pppp} \expval{qp|ps}  + 2(1-\delta_{pr})\Gamma_{prpr}\expval{qr|ps}
\\
&+(1-\delta_{pr}) \Gamma_{prrp}(\expval{qp|rs} + \expval{qr|sp}).
\end{split}
\label{eq:B3}
\end{equation}
Exploiting the relationships $\Gamma_{pmmp} = \Gamma_{pmpm}$ when $p=m$, 
$\delta_{pr} \Gamma_{prpr}\expval{qr|ps} = \Gamma_{pppp}\expval{qp|ps}$, 
and $(1-\delta_{pr})\delta_{pr} = 0$ gives
\begin{equation}
\begin{split}
Y_{pqrs} &=\delta_{pr}  \sum_{m}\Gamma_{pmpm}\expval{qm|sn} 
\\
&+\delta_{pr}  \sum_{m}(\Gamma_{pmmp} - \delta_{pm}\Gamma_{pmpm})\expval{qm|ms}
\\
&+2\Gamma_{prpr} \expval{qr|ps}
\\
&+(\Gamma_{prrp} - \delta_{pr}\Gamma_{prpr})(\expval{qp|rs} + \expval{qr|sp}),
\end{split}
\end{equation}
where the $(1-\delta_{pr})$ term on the last line of Eq.~\eqref{eq:B3} is dropped by noting that 
$(\Gamma_{prrp} - \delta_{pr}\Gamma_{prpr}) = 0$ when $p=r$.
The coupling constants $a_{pq} = \Gamma_{pqpq}$ and $b_{pq} = \Gamma_{pqqp} - \delta_{pq}\Gamma_{pqpq}$
can now be inserted to obtain
\begin{equation}
\begin{split}
Y_{pqrs} &=  \delta_{pr} \sum_{m} (a_{pm}\expval{qm|sm} + b_{pm}\expval{qm|ms} )
\\
&+  2a_{pr} \expval{qr|ps} + b_{pr}(\expval{qp|rs} + \expval{qr|sp}).
\end{split}
\end{equation}
Using the definition of the Coulomb and exchange matrices in Eq.~\eqref{eq:JKmo} 
allows this expression to be given as
\begin{equation}
\begin{split}
Y_{pqrs} &=  \delta_{pr} \qty(n_\cP J_{qs} + K^{\cP}_{qs} )
\\
&+2a_{pr} \expval{qr|ps} + b_{pr}(\expval{qp|rs} + \expval{qr|sp}).
\end{split}
\end{equation}
The definition of the generalized Fock matrix elements in Eq.~\eqref{eq:genfock_JK} then yields
\begin{equation}
\begin{split}
Y_{pqrs} &= \delta_{pr} \qty(F^\cP_{qs} - n_p h_{qs}) 
\\
&+ 2a_{pr}\expval{qr|ps} + b_{pr} \qty(\expval{qp|rs} + \expval{qr|sp}).
\end{split}
\label{eq:Ypqrs}
\end{equation}
Finally, substituting Eq.~\eqref{eq:Ypqrs} into Eq.~\eqref{eq:hessian} and using 
$\gamma_{pr} = \delta_{pr} n_\cP$ yields the second derivatives in Eq.~\eqref{eq:Qpqrs}. 

\section{Optimal ordering of open-shell orbitals}
\label{sec:optimal_ordering}

When initialising a CSF using localized orbitals, it is important that the orbitals are 
assigned to different open shells in a physically meaningful way. 
For example, in an antiferromagnetic bimetallic complex, the unpaired electrons on each metal 
centre should occupy a shell with local ferromagnetic coupling, while the long range
coupling between the two metal centres should be antiferromagnetic. 
One approach to achieve the optimal assignment of initial local orbitals to each shell is 
use combinatorial optimization to find the ordering that minimizes the exchange energy.
While this task has previously been addressed using simulated annealing in Ref.~\citenum{Dobrautz2022},
here I define a simple local search based on swapping two spatial orbitals at a time.

The starting point for this optimization is to note that the exchange energy for a CSF with 
$\No$ open-shell orbitals ordered as $\{\MO_1, \MO_2, \cdots, \MO_{\No} \}$ is given by\cite{Dobrautz2022} 
\begin{equation}
E_{\text{x}}^{\text{os}}(\bm{\mu}) = \sum_{w>v} b_{vw} K_{vw} (\bm{\mu})
\end{equation}
where the exchange integrals $K_{vw} (\bm{\mu})=\expval{vw|wv}$ depend on the ordering $\bm{\mu}$ 
of the spatial orbitals, e.g.\ $\bm{\mu} = $(1-2-$\dots$-$\No$).
Here, the indices $v,w$ correspond to the orbital at a given position in $\bm{\mu}$.
Only the energy contribution $E_{\text{x}}^{\text{os}}(\bm{\mu})$ is affected when the ordering 
of open-shell orbitals changes.
Local steps in the combinatorial space of orbital 
orderings can be defined by permuting two spatial orbitals.
For example, applying the permutation (12) to 
$\bm{\mu} = $(1-2-$\dots$-$\No$) gives $(12)\bm{\mu} = $(2-1-$\dots$-$\No$), 
which changes the energy if the orbitals in the first and second positions occupy different shells.
The energy difference $\Delta E_{\text{x}}^{\text{os}}$ for two orbital orderings 
$\bm{\mu}$ and $\bm{\mu}'$ can be computed as
\begin{equation}
\Delta E_{\text{x}}^{\text{os}} = \sum_{w>v} b_{vw} \qty(K_{vw} (\bm{\mu}') - K_{vw} (\bm{\mu}))
\end{equation}
Starting from a particular ordering $\bm{\mu}_0$, a local downhill search can then be defined by systematically 
testing the change in $E_{\text{x}}^{\text{os}}(\bm{\mu})$ when two orbitals are swapped and accepting 
the swap if it lowers the energy.
In practice, all possible swaps are considered sequentially, and the process is repeated until 
no further energy reduction can be achieved.
While this approach is not guaranteed to find the globally optimal ordering of orbitals, 
it can significantly improve the initial guess for CSF optimization.

\section*{References}
\bibliography{manuscript}

\cleardoublepage
\begin{figure*}
{\large \sf TOC Graphic}\\
\fbox{\includegraphics[width=0.5\linewidth]{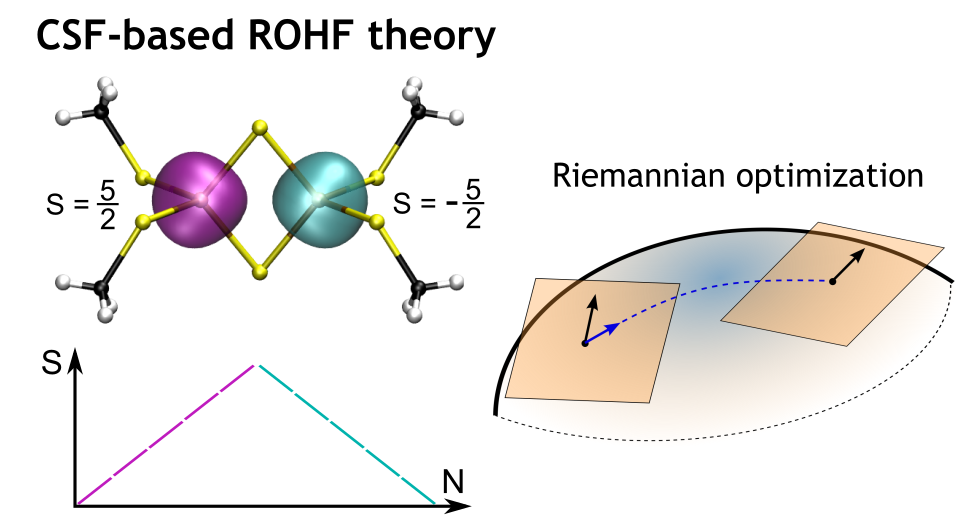}}
\end{figure*}
\end{document}